# Skyrmionics in correlated oxides

Zhi Shiuh Lim,[1][*] Hariom Jani,[1][*] T. Venkatesan,[2,3] and A. Ariando[1][†]

[1] Department of Physics, National University of Singapore, Singapore;
[2] Department of Electrical & Computer Engineering, National University of Singapore, Singapore;
[3] Center for Quantum Research and Technology, University of Oklahoma, USA;

[*]Authors have contributed equally

[†]ariando@nus.edu.sg

**Abstract**

While chiral magnets, metal-based magnetic multilayers or Heusler compounds have been considered as the material workhorses in the field of skyrmionics, oxides are now emerging as promising alternatives, as they host special correlations between the spin–orbital–charge–lattice degrees of freedom and/or coupled ferroic order parameters. These interactions open new possibilities for practically exploiting skyrmionics. In this article, we review the recent advances in the observation and control of topological spin textures in various oxide systems. We start with the discovery of skyrmions and related quasi-particles in bulk and heterostructure ferromagnetic oxides. Next, we emphasize the shortcomings of implementing ferromagnetic textures, which have led to the recent explorations of ferrimagnetic and antiferromagnetic oxide counterparts, with higher Curie temperatures, stray-field immunity, low Gilbert damping, ultrafast magnetic dynamics, and/or absence of skyrmion deflection. Then, we highlight the development of novel pathways to control the stability, motion and detection of topological textures using electric fields and currents. Finally, we present the outstanding challenges that need to be overcome to achieve all-electrical, non-volatile, low-power oxide skyrmionic devices.

**Keywords:** Spintronics, Topology, Oxides, Antiferromagnets, Skyrmions, Beyond-Moore Computing

# 1. Introduction

Skyrmionics is a burgeoning field in spintronics that aims to exploit real-space whirling *topological magnetic textures* as mobile information bits ('0' and '1') in beyond-Moore computing paradigms.[1] Herein, textures such as skyrmions, bimerons, merons and others, illustrated in **Figure 1**, are composed of spins winding in a specific sense endowing them *topological protection*, and allowing them to effectively behave like quasi-particles that can be controlled and processed electrically. Skyrmionic topological racetrack memory, akin to magnetic tunnel junction based random access memory,[2] brings the benefits of fast access and eliminates mechanical moving parts, markedly improving upon the state-of-the-art hard disk drives. Moreover, in the bit control process, in contrast to regular ferromagnetic domain switching or domain wall (DW) motion driven by spin torques,[3] skyrmionic platforms allow the reduction of threshold current density ($J_C$) by up to five orders-of-magnitude, afforded by skyrmions evading defect pinning because of their topological protection.[4-7] Due to these unique advantages, skyrmionics could open a promising pathway to build *fast*, *dense* and *energy-efficient* non-volatile computing.

Various chiral topological textures can be classified by their unique topological charge, helicity, and orientation of background spins, see **Table 1**. The topological charge ($Q_{\hat{n}}$) quantifies the integrated solid angle spanned by the spatially varying local order parameter $\hat{n}$,

$$Q_{\hat{n}} = \frac{1}{4\pi} \iint d^2\mathbf{r} \ \hat{n} \cdot \left[\frac{\partial \hat{n}}{\partial x} \times \frac{\partial \hat{n}}{\partial y}\right]$$

which corresponds to the magnetization $\hat{m}$ in ferromagnetic (FM) or ferrimagnetic (FiM) systems and the Néel vector $\hat{l}$ in the antiferromagnetic (AFM) counterparts. Topological textures are stabilized by a negative DW energy,[8, 9] that can be achieved by the Dzyaloshinskii–Moriya Interaction (DMI) involving a broken spatial inversion symmetry associated with the $D_n$, T or O (for Bloch skrymions), $C_{nv}$ (for Néel skyrmions) and $D_{2d}$ (for antiskyrmions) point group symmetries, or through geometrical frustration.[10-12] In the absence of a DW energy-lowering agent, the dipolar /demagnetization field may also stabilize multichiral magnetic textures. Most recently, the Kibble–Zurek mechanism,[13, 14] which exploits magnetic phase transitions to nucleate multichiral topological textures, has also joined this list.

Figure 1 illustrates various magnetic textures found thus far, albeit non-exhaustively, and indicates the corresponding stabilizing mechanisms. Typically, nucleation of these textures is assisted by thermal excitations, usually not found at the ground state, when the competing terms of exchange, anisotropy, DMI/frustration, and external magnetic field are present in an effective magnetic Hamiltonian. With increasing magnetic field, several phase transitions can be seen from stripes, to skyrmion lattice (SkL) before saturating into a trivial collinear magnetic state.[15] By tuning the strength of DMI relative to exchange and anisotropy it is possible to go from thermodynamically stable topological lattices or spin spirals in the strong DMI regime above a critical threshold, to isolated metastable topological quasiparticles in the intermediate/weaker DMI regime. The latter scenario is practically more relevant as it enables these textures to serve as individually addressable bits in racetrack implementations.[16, 17]

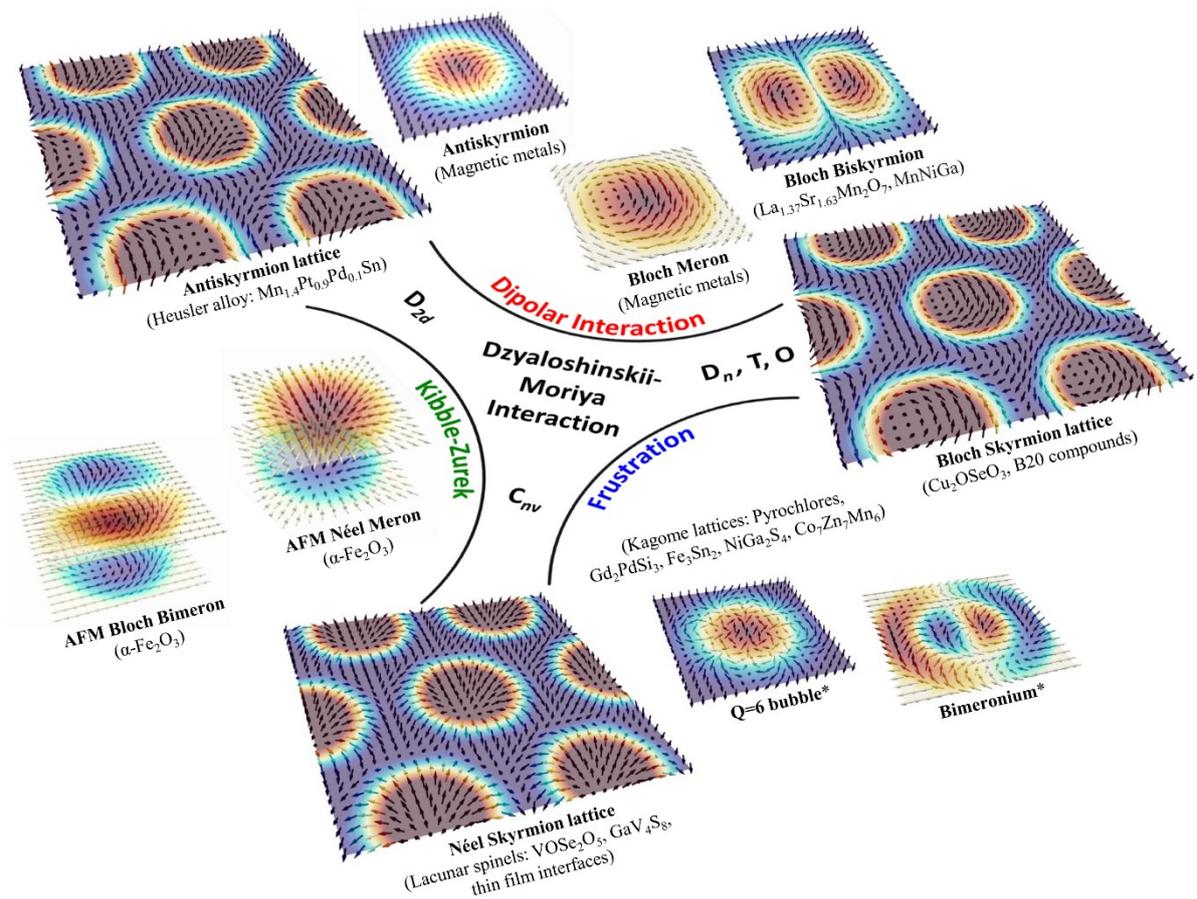

**Figure 1.** Summary of mechanisms corresponding to various magnetic textures found to date, with possible overlaps. $C_{nv}$, $D_n$, $T$, $O$ and $D_{2d}$ are Schoenflies notations of point group symmetries. Realistic material examples are given in the brackets.

**Table 1**: Summary of various topological textures and their structural properties. Sign of the topological charge depends on the spin orientation at the cores, such that a quasiparticle and its antiparticle would have opposite topological charges for the same core orientation. Helicity refers to the sense of spin twisting in the textures.

| Texture | $Q_{\hat{m}}$ | $Q_{\hat{l}}$ | Helicity | Background |
|---|---|---|---|---|
| Bloch Skyrmion | $\pm 1$ | - | Helicoidal | Out-of-plane |
| Néel Skyrmion | $\pm 1$ | - | Cycloidal | Out-of-plane |
| Antiskyrmion | $\mp 1$ | - | Mixed sectors | Out-of-plane |
| Bloch Meron | $\pm 1/2$ | - | Helicoidal | In-plane |
| Néel Meron | $\pm 1/2$ | - | Cycloidal | In-plane |
| Antimeron | $\mp 1/2$ | - | Mixed sectors | In-plane |
| Bloch Bimeron | $\pm 1$ | - | Helicoidal | In-plane |
| Néel Bimeronium | $0$ | - | Cycloidal | In-plane |
| Bloch Biskyrmion | $\pm 2$ | - | Helicoidal | Out-of-plane |
| AFM Néel Skyrmion | $0$ | $\pm 1$ | Cycloidal | Out-of-plane |
| AFM Néel Meron | $0$ | $\pm 1/2$ | Cycloidal | In-plane |
| AFM Bloch Meron | $0$ | $\pm 1/2$ | Helicoidal | In-plane |
| AFM Antimeron | $0$ | $\mp 1/2$ | Mixed sectors | In-plane |
| AFM Bloch Bimeron | $0$ | $\pm 1$ | Helicoidal | In-plane |

## 1.1 Imaging and Detection techniques

To date, significant advances have been made in skyrmionics thanks to a plethora of microscopic imaging and detection techniques, which harness neutrons, photons, electrons or stray magnetic fields. This includes neutron scattering,[18] magneto-optical Kerr effect (MOKE),[19] magnetic force microscopy (MFM),[17] spin-polarized scanning tunnelling microscopy (Sp-STM),[20] Lorentz transmission (LTEM),[21] and dichroic photoemission electron microscopy (PEEM),[22] scanning transmission X-ray microscopy (STXM),[23] nitrogen-vacancy (NV) magnetometry.[24] Among these, X-ray techniques offer the unique advantage of resonant *element-specific imaging*, thereby, providing an unambiguous picture of the underlying textures. Furthermore, electrical detection via the anomalous topological contributions to the Hall effect (THE) has also served as a promising technique.[25] However, THE only provides a tell-tale sign of the existence of chiral textures, without definitively confirming if they are topological quasiparticles. Lastly, the tunnelling non-collinear magnetoresistance (TNcMR) effect, where the tunnelling current across a multi-layer junction depends on the non-collinearity of spins in the racetrack layer, also promises electrical detection.[26] Future development of skyrmion racetrack memory is expected to employ some form of TNcMR for all-electrical readout of topological textures.

**1.2 Why Correlated Oxides?**

To harness topological textures practically, it is important to develop materials that can stabilize these textures across a wide range of temperatures including room temperature, and develop all-electrical pathways to nucleate, move, read and destroy them in patterned tracks.

A large family of topological textures and control mechanisms have been reported in magnetic metal-based heterostructures, or chiral magnets such as Heusler compounds, and B20 systems,[1] making them favourites for developing skyrmionics. However, correlated oxide magnets have also drawn considerable interest as they offer the following distinctive opportunities. (i) They host coupled ferroic order parameters (e.g., magneto-electric or multiferroic)[27, 28] or special correlations between various internal degrees of freedom (e.g., charge, spin, orbital and lattice), which are quite susceptible to external perturbations. These couplings open novel practical pathways to control both intrinsic and emergent magnetic properties, be it anisotropy, exchange or DMI. (ii) Oxides often have lower and tunable charge carrier density making them quite susceptible to electric fields, ideal for non-volatile control. (iii) Furthermore, high quality crystalline oxide films with precisely controlled interfaces/terminations, and heterostructures thereof, hosting very low defects can be fabricated via standard growth techniques. This could help flatten the pinning landscape and reduce threshold current densities required to move textures, ideal for energy-efficient control. These aspects have primed oxides to emerge as potential hosts to build electrically controlled, non-volatile skyrmionic devices.

This review is dedicated to the two overarching thrusts in the field of oxide skyrmionics – i.e., the development of *materials* and *control mechanisms* – that could help to pave the way toward practical oxide devices.

**2. Materials development**

The quest for magnetic textures in oxides began as early as the 1970s, in orthoferrites,[29] garnet-ferrites,[30, 31] and hexaferrites[32, 33] during the bubble memory era, and more recently in the 2010s in manganites[34-39] where the materials were prepared as thin plates to satisfy the stability conditions imposed by dipolar interaction. Imaging was performed by magneto-optical Microscopy, and more recently by LTEM. Hence various textures, including trivial bubbles,[38, 39] (counter)clockwise Bloch skyrmions[38] and compound textures such as biskyrmions[36] or complex multi-ring skyrmions[33] were found to exist across different parts of the parameter space. Many of these textures are illustrated in Figure 1.

$La_{1.37}Sr_{1.63}Mn_2O_7$ was one of the celebrated examples that hosted a dense biskyrmion lattice (**Figure 2**a–b) and recorded promisingly low threshold $J_C$ of ~$10^7$ A/m$^2$ for their electrically driven motion. Almost concurrently, bulk $Cu_2OSeO_3$ crystal was found to host Bloch-type SkL as it possesses bulk DMI similar to B20 compounds, due to the *T* symmetry.[40] In contrast, the lacunar spinel $VOSe_2O_5$ was found to stabilize Néel-type SkL and incommensurate (IC) spin-crystal neighbouring phases as verified by neutron scattering.[41] Its polar crystal structure was found to play the crucial role of providing the $C_{4v}$ symmetry, although no sharp material interface was involved. These discoveries in bulk systems, inspired further investigations into oxide thin films.

## 2.1 Ferromagnetic textures

Over the last few decades, various layer-by-layer or step-flow oxide growth techniques have emerged which have enabled the creation of high quality interfaces for novel heterostructure design. Such capability has opened doors for stabilizing Néel-type skyrmions by engineering interfacial DMI. For instance, in ferromagnetic $SrRuO_3$ thin films, $Ru^{4+}$ 4d $t_{2g}^4 e_g^0$ electronic configuration is known to endow several interesting properties, such as the minority-spin double exchange and the unquenched orbital angular momentum (L = 1) – allowing large spin–orbit coupling (SOC) and perpendicular magnetic anisotropy (PMA). The partially filled $t_{2g}$ orbitals also form bands with various *k*-space Chern numbers,[42, 43] producing a large anomalous Hall effect (AHE). Subsequently, via first-principles calculations and half-integer x-ray Bragg diffraction,[44] the monoclinic and tetragonal phases of $SrRuO_3$ can be distinguished especially in the sign of dominant AHE conductivity. On the other hand, in semi-metallic $SrIrO_3$, the $Ir^{4+}$ 5d $J_{eff}$ = 1/2 band near Fermi level ($E_F$) is also understood to have Dirac-like band dispersion due to its strong SOC.[45] Hence, (001)-oriented $SrIrO_3$/$SrRuO_3$ interfaces were explored for topological textures where strong DMI could be expected due to interfacial inversion symmetry breaking. Indeed, THE signals were found in the ultrathin, low Curie temperature regime hinting at the existence of magnetic skyrmions.[46] This THE signature, usually appearing as hump-shaped anomalies in the Hall transport measurements, emerges from the deflection of mobile spin-polarized electrons as they accumulate a Berry phase while traversing the emergent magnetic field from the non-coplanar magnetic moments constituting the topological textures.[25, 47] This Berry phase, which is a geometric phase accumulated in the wavefunction of the mobile electrons, results from the sizable exchange interaction between the mobile spins and the spatially varying local moments. Subsequently, similar THE-like signals were also discovered in single-layer ultrathin $SrRuO_3$[48] and ferroelectric/$SrRuO_3$ heterostructures.[49]

However, an alternative interpretation of THE soon arose, where the hump-shaped Hall features were reproduced by two overlapping AHE loops, with Langevin function shapes of opposite signs.[43, 50] Here, AHE refers to the Karplus–Luttinger intrinsic mechanism that must involve SOC directly in driving electron deflection,[51] unlike the geometrical THE where a scalar moment chirality is needed, but not SOC. This implied the presence of trivial magnetic domain inhomogeneities, most likely a coexistence of monoclinic and tetragonal phases of $SrRuO_3$, and hence bringing into question the existence of genuine topologically protected magnetic skyrmions. In particular, magnetic states in the aforementioned $SrIrO_3$/$SrRuO_3$ bilayers[46, 52] and the atomic terrace-engineered ultrathin $SrRuO_3$ films[53] were found to be distinct, when measured by low-temperature MFM. Skyrmion-like bubbles were imaged in the former cases[52] (Figure 2c–d), while the latter showed straight domains segregated along the atomic terraces,[53] albeit hump-shaped THE features were present in both systems. Hence, the debate around THE has demanded further support from high-resolution magnetic imaging to unambiguously confirm the topological origin for each case.

More recently, MFM imaging of $SrRuO_3$/$PbTiO_3$(001) with Fourier transform revealed a "square-lattice IC-phase" formed by two perpendicularly propagating cycloids.[54] Such IC-phase is similar to those in bulk $GaV_4S_8$ and $VOSe_2O_5$, also with $C_{3,4v}$ symmetries.[41, 55] Hence, the mentioned Hall humps were concluded as true THE due to the scalar moment chirality present in the IC-phase.

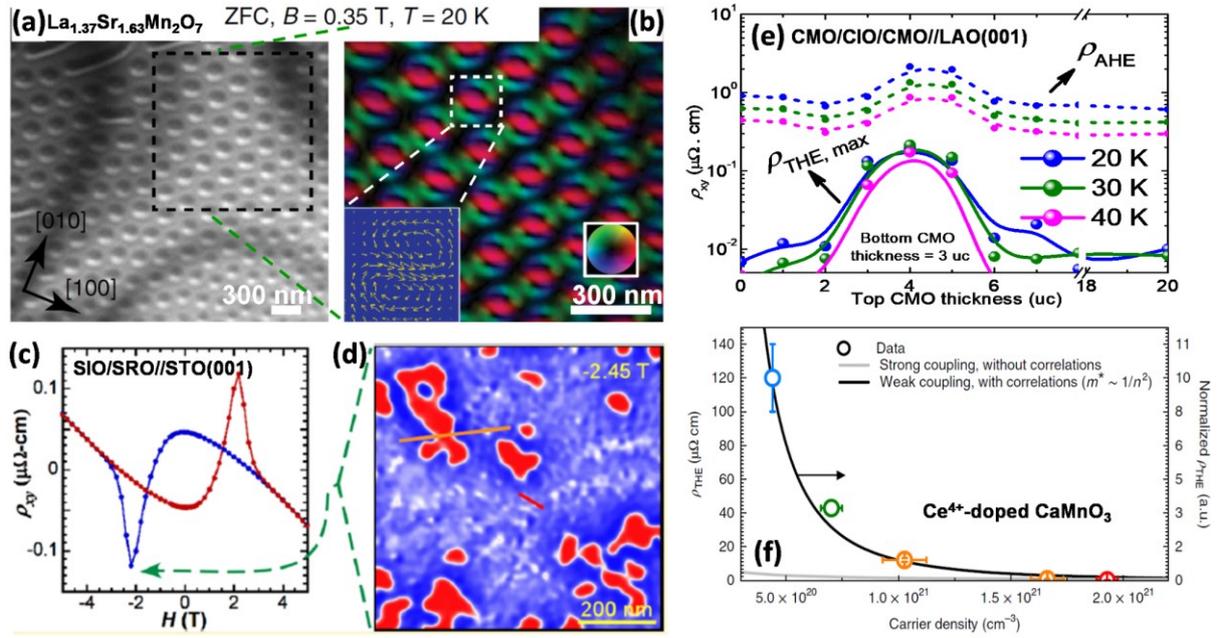

**Figure 2.** Realization of FM topological textures. (a, b) Biskyrmion lattice imaged from an La$_{1.37}$Sr$_{1.63}$Mn$_2$O$_7$ bulk crystal via LTEM, with the magnetic orientation colour wheel, scale bar and local magnetic moment vectors shown as insets.[36] Hump-shaped THE features (c) and MFM image of isolated skyrmionic bubbles (d) observed in SIO/SRO bilayers.[52] (e) Sensitive dependence of THE on CaMnO$_3$ thickness in the CMO/CIO/CMO trilayer structure.[56] (f) CCMO thin film showing correlation enhanced THE versus carrier density.[57]

Meanwhile, the interfacial 3$d$–5$d$ cationic charge-transfer scheme[58] became an insightful concept for achieving t$_{2g}$-dominated and topologically nontrivial valence band in manganites. The large PMA and AHE in SrMnO$_3$/SrIrO$_3$[59, 60] and CaMnO$_3$/CaRuO$_3$[61] superlattices are very similar, and also resemble the double-perovskites A$_2$B′B″O$_6$ studied earlier[62] where the inter-B-site super-exchange and ferro-/ferrimagnetism is absent when the parent materials are separated. Furthermore, the incorporation of 5$d$ Ir$^{4+}$ at sharp interfaces opens the possibility of creating large SOC and interfacial DMI. Subsequently, as an effort to extend skyrmion search into antiferromagnets, obvious THE signals were found in LaMnO$_3$/SrIrO$_3$ superlattices[63] and CaMnO$_3$/CaIrO$_3$/CaMnO$_3$ trilayer (Figure 2e).[56] respectively. In the former system, the distinct cationic oxidation states (La$^{3+}$Mn$^{3+}$ and Sr$^{2+}$Ir$^{4+}$) were discussed as the source of interfacial broken inversion symmetry. While in the latter, random stacking faults were found at the bottom CaIrO$_3$/CaMnO$_3$ interface of the trilayer for avoiding DMI cancellation. On the other hand, the lightly Ce$^{4+}$-doped CaMnO$_3$ single-layer thin film was realized as a good candidate for stabilizing magnetic bubbles by dipolar interaction,[57] confirmed by low-temperature MFM. Here, a giant THE signal was reported that diverged with reducing carrier density, which is consistent with a correlation-enhanced THE model (Figure 2f).[64]

While FM topological textures have significant promise for skyrmionics, they suffer from some critical drawbacks. They typically require external magnetic fields for their stabilization.[49, 57] Next, they exhibit undesired sideway deflection when driven by lateral currents (known as skyrmion Hall effect).[19, 23] Moreover, the long range dipolar interactions in FMs further restricts skyrmion size scaling.[65] These limitations can be overcome by constructing topological counterparts in systems with multiple magnetic sublattices, which partially/fully compensate each other, such as in FiM/AFM materials.[65, 66] These systems have the following exceptional properties: (i) Magnetic compensation weakens internal dipolar fields and makes these materials impervious to stray fields. This allows the local AFM order to be *robust*, while opening up the *densification* of texture sizes. (ii) Many FiM/AFM oxides have high ordering temperatures (above 300 K) which assist in *widening the thermal stability* window of textures.[22, 67] (iii) Presence of the exchange amplification effect combined with very low magnetic Gilbert damping ($10^{-4}$–$10^{-2}$), observed in insulating oxides,[68, 69] unlocks ultrafast magnetic dynamics[70] and relativistic motion[71][72] of spin textures (~1-10 km/s). This is expected to be about 1-3 orders of magnitude faster than FM counterparts, under similar electrical excitation, enabling *high speed* and *energy efficient* processing. (iv) Finally, specific AFM oxides also naturally host coupled magnetic and electrical degrees of freedom, promising electrical control of magnetic textures.[73, 74] These aspects have led to a recent surge of investigations into FiM/AFM topological textures.

**2.2 Ferrimagnetic textures**

A tell-tale sign for the presence of chiral textures in FiM oxide films was initially obtained by Hall transport, where ultrathin strained samples of rare-earth garnets ($Tm_3Fe_5O_{12}$) interfaced with a heavy-metal overlayer[67, 75, 76] exhibited sizable THE cusps. These signals appeared near and above room temperature, coinciding with the significant weakening of the out-of-plane anisotropy.[67] It was suggested that since garnets are insulators these THE signals indirectly probe chiral spin textures through currents in the heavy-metal layer, either via proximity-induced magnetism[77] and/or spin-Hall THE (SH-THE).[75] However, the origin of DMI in typical garnet heterostructures has been an actively debated subject. While some reports underscore the role of the heavy-metal[67, 75, 76] or substrates,[78, 79] detailed studies have revealed that DMI is expected to principally originate from the strong intrinsic spin–orbit coupling of the rare-earth ions present near inversion symmetry breaking interfaces, which is enhanced below room temperature.[79] Hence, it is important to directly image topological FiM textures to delineate the origin of THE.[67, 75, 76] To this end, the presence of bubble-like topological textures in $Tm_3Fe_5O_{12}$ films, both with and without heavy-metal overlayers, has been confirmed via Kerr microscopy[78] and more exhaustively by STXM and PEEM,[80] **Figure 3**a,b. Interestingly, it was shown that both electrical currents and ultrafast light pulses can be harnessed to generate and control the bubbles at room temperature,[80] which is promising for practical exploitation. However, it remains to be verified if the evolution of these bubbles is directly linked to aforementioned THE signals, by performing electrical and imaging experiments in tandem.

## 2.3 Antiferromagnetic textures

While textures in FM and FiM materials can be visualized and controlled readily by standard magnetic techniques, it is difficult to directly translate these approaches to AFM counterparts due to the compensation of constituent magnetic sublattices. The first problem of imaging the AFM order parameter (i.e. the Néel vector), can be tackled via synchrotron-based linear dichroic X-ray imaging,[81] where combining angle-dependent images has been used to generate Néel vector phase-maps.[82, 83] However, the second problem of generating AFM topological textures has been a hard nut to crack, as AFMs respond poorly to external magnetic field – one of the standard ingredients for generating FM skyrmions. To overcome this, inspiration was drawn from a half-century-old idea of cosmic string formation, called the Kibble–Zurek mechanism,[13, 14] where a phase transition was suggested to generate topological defects under specific constraints, for instance U(1)-like symmetry.[82]

Such a constraint is approximately satisfied in the earth-abundant layered-AFM oxide – α-$Fe_2O_3$ – across the Morin transition temperature ($T_M$) where the Néel vectors reorient from out-of-plane to in-plane. This approach has enabled the recent breakthrough[22] of reversible and field-free generation of a family of topological AFM textures, including Néel and Bloch AFM merons, antimerons, and bimerons, Figure 1 and 3c, that are stable in the in-plane state at and above room temperatures. Here, (anti)merons and bimerons are equivalent to half-skyrmions and in-plane-rotated skyrmions, respectively, with their core dimensions directly tunable through temperature,[22] chemical/ionic doping,[84] strain,[85] or heterostructure design.[82, 86] In the latter case, coupling α-$Fe_2O_3$ with a Co FM layer resulted in flat AFM (anti)vortices coupled to FM (anti)merons.[82, 86] Interestingly, simulations suggest that introducing sufficient interfacial DMI along with carefully tuned exchange ($A$) and anisotropy ($K$) energy, could result in the stabilization of homochiral topological textures, including not only Néel AFM merons and bimerons above $T_M$ (in-plane state ), but also the hitherto unreported Néel AFM skyrmions below $T_M$ (out-of-plane state ). The ability to use the Kibble–Zurek transition significantly widens the scope of materials that could be exploited for AFM topological spintronics, as a host of AFM oxides (e.g., orthoferrites, layered ferrates, orthochromites) also possess first and/or second-order Morin transitions.[70]

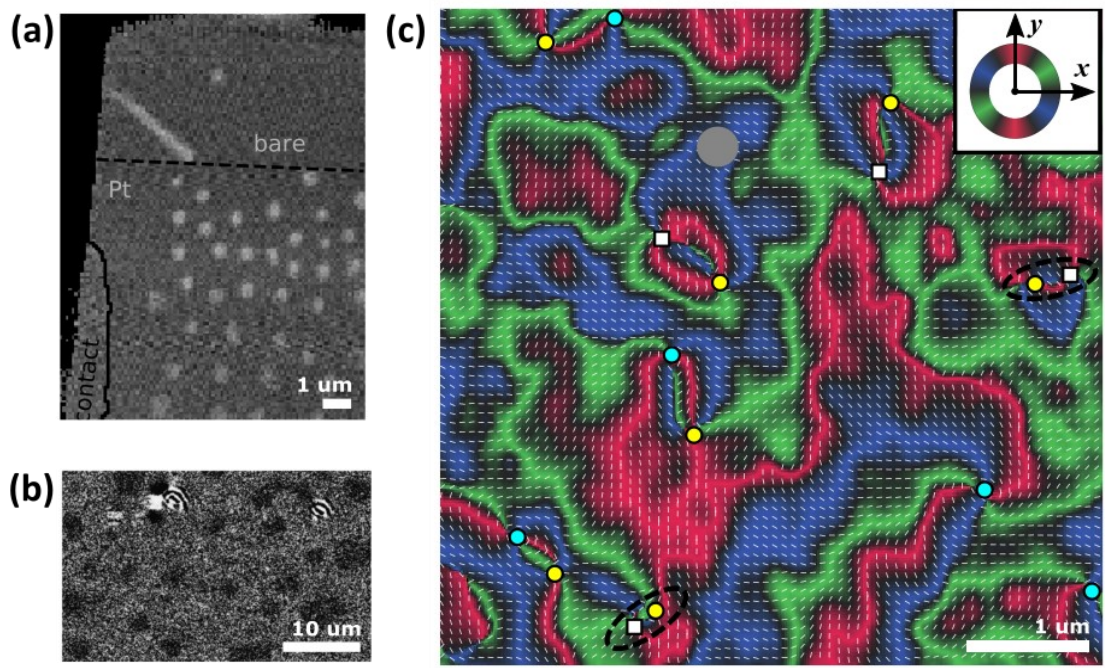

**Figure 3.** Realization of FiM and AFM topological textures. (a, b) Observation of FiM bubbles, seen as small circles with different contrast relative to the background, in $Tm_3Fe_5O_{12}$ films measured via (a) STXM[80] and (b) MOKE.[78] (c) Formation of AFM Bloch merons (blue circles), Néel merons (white squares), antimerons (yellow circles), bimerons or trivial meron pairs (pairs encircled in dashed black ellipses) in α-$Fe_2O_3$, imaged via angle-dependent PEEM.[22] Red-Green-Blue contrast and the white dashes depict the in-plane AFM order. All measurements (a-c) were performed at room temperature.

Another material of interest is the AFM oxide α-$Cr_2O_3$, which was found to exhibit THE signals near and above the Néel ordering temperature, when its thin films were interfaced with a Pt over-layer.[87] As α-$Cr_2O_3$ is magnetoelectric,[73] this could open an interesting pathway to electrically control AFM topological textures. However, it must be noted that while SH-THE assisted transport signals hint at the presence of chiral spin textures, further imaging studies in tandem are required to unequivocally confirm their topological origin.

Finally, in analogy with synthetic antiferromagnets (SAF) made of metallic multilayers using Ruderman–Kittel–Kasuya–Yosida interaction,[88] oxide-SAF have been achieved using the interlayer-exchange coupling in $La_{0.7}Ca_{0.3}MnO_3/CaRu_{0.5}Ti_{0.5}O_3$ superlattice, where the $Mn^{4+}/Ru^{4+}$ coupling is antiferromagnetic.[89] By controlling the superlattice repetitions, the magnetization at zero-field can be tuned between FM/AFM orientations. Hence, this could be also become a potential system for exploring SAF-based topological textures.

## 3. Control mechanisms

To enable practical topological racetracks, three crucial prerequisites need to be met: (i) control over the stability landscape for texture generation/annihilation, (ii) control of texture motion, and (iii) the readout mechanism that detects topological textures. These features have to be integrated into an *all-electrical platform*. To this end, oxides offer several advantages not found in other materials.

### 3.1 Controlling the stability landscape of topological textures

To precisely generate/annihilate skyrmions with low power consumption, a zero-current, magnetic-field-free, non-volatile electric field controlled magnetic landscape is highly desirable. This can be achieved either with systems possessing coupled ferroic orders, either intrinsically or via interfacial interactions. In metallic skyrmion platforms, the state-of-the-art approaches involve skyrmion nucleation via current-based spin-orbit torques [90] or the voltage-controlled magnetic anisotropy (VCMA) by carrier density modulation.[91] Hence, coupling to the ferroelectric order in oxides offers a new advantageous scheme providing a non-volatile field control.

Among oxide bulk crystals, $Cu_2OSeO_3$ is a celebrated compound for its multiferroic functionality. Besides housing a Bloch-type SkL, its *d–p* hybridization allows magnetoelectric coupling between the electric polarization and magnetic moment.[92] Using alternating field (AC) magnetometry with an applied magnetic field $H_{[111]}$ to obtain a SkL, the magnetoelectric coupling was observed by a depression in the real-part of AC-susceptibility with concomitant enhancement in polarization $P_{[111]}$.[93, 94] Subsequently, numerous other characterizations for multiferroic SkL followed. Particularly, using microwave resonance of the skyrmion rotation and breathing modes, a non-reciprocal directional dichroism effect was observed.[95] Besides, a magnetoelectric cooling process was demonstrated (**Figure 4**a–b), where the skyrmion phase space is systematically enlarged by cooling with an electric field. A sweeping electric field also produced hysteresis in the AC-susceptibility, as an evidence of tuning the magnetic phases.[96] It has been shown that preparing thin plates can enlarge the skyrmion phase-space.[93] However, using $Cu_2OSeO_3$ for practical developments has remained challenging due to the low Curie temperature and difficulty of thin-film growth. Hence, it would be interesting to search for topological textures in room temperature multiferroic or magneto-electric systems than can be readily grown via standard approaches.

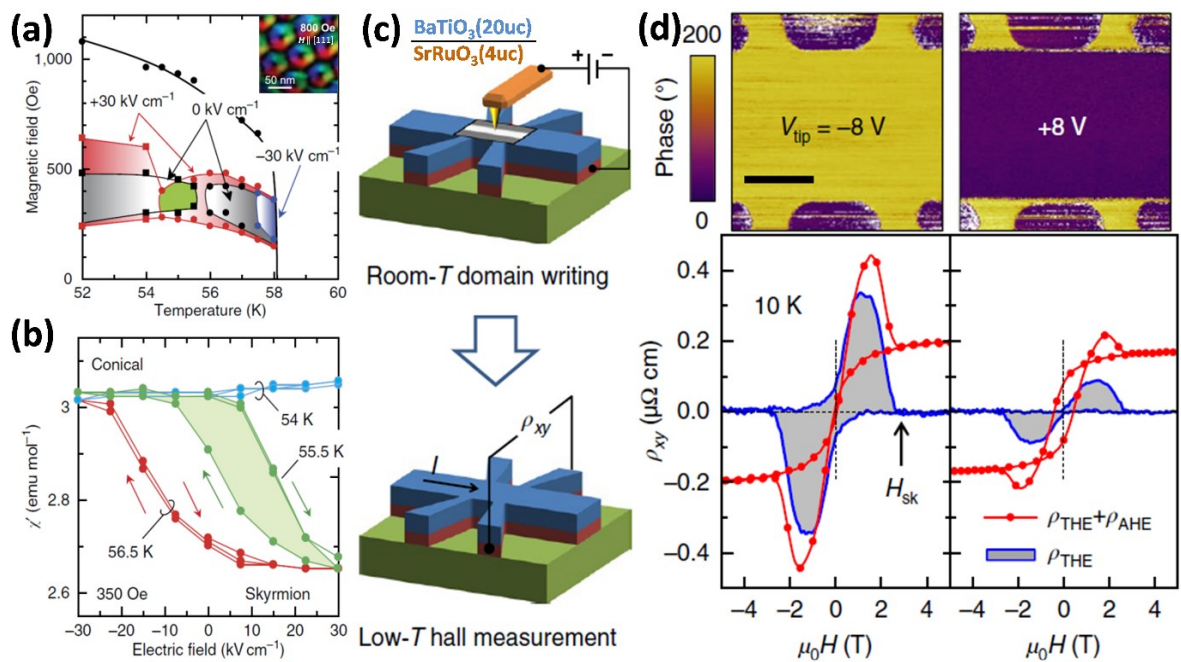

**Figure 4.** Observation of electrical control of FM topological textures. (a) Tuning skyrmion stability in $Cu_2OSeO_3$ by applying electric fields,[96] with the image of SkL in the inset adapted from.[93] Green region represents the window in which non-volatile switching of the skyrmion phase is possible using electric fields. (b) Hysteresis of AC-susceptibility, representing switching between conical and skyrmion states, with varying electric field in the overlapping region shown in (a). (c) FE domain writing and measurement schemes in Ref[49] and (d) THE, and thereby, skyrmion density modulations corresponding to percentage of switched FE domains within the Hall bar channel.

In oxide heterostructures such as $SrRuO_3$/titanate ferroelectric (FE) interfaces in (001)-orientation, the tetragonal distortion has been discussed to propagate into a thin layer of $SrRuO_3$ resulting in an interfacial DMI with $C_{4v}$ symmetry for stabilizing Néel-type skyrmions. In a recent breakthrough exploiting ultrathin BaTiO$_3$/SrRuO$_3$ heterostructures,[49] electric field applied through a sharp tip was used to locally switch the BaTiO$_3$ polarization at room temperature in a Hall bar channel, which tuned the interfacial DMI and thereby the topological properties, Figure 4c. Changes to THE signals and MFM images were measured at low temperatures, and correlated to the percentage of switched FE domains (Figure 4c–d). However, several caveats can be noted: with the bottom SrRuO$_3$ being obscured by the top BaTiO$_3$, quantifying the changes in DMI by MFM is difficult due to the weaker signal. Besides, BaTiO$_3$ is known to undergo several structural phase transitions upon cooling,[97] which could result in a change of the FE polarization easy-axis. Hence, it is preferable to perform the electrical switching and magnetic imaging experiments isothermally to make unambiguous correlations.

These developments, harnessing the ferroelectric order in natural/artificial multiferroic oxides, have opened non-volatile pathways to tune topological stability. Next, it will be crucial to realize this electric field control over topological textures in FiM/AFM materials at room temperature.

## 3.2 Moving and detecting topological textures

For moving topological textures, three underlying mechanisms have been identified, namely the thermal effect, spin-transfer torque (STT) and spin-orbit torque (SOT).

Firstly, in thin Cu$_2$OSeO$_3$ plates, a local temperature imbalance was found to produce magnon current components propagating parallel and perpendicular to the temperature gradient. These magnon currents are deflected by the emergent magnetic field of the topological SkL due to spin torques, and in return the skyrmions experience a reaction force resulting in their motion. Radial thermal gradients have produced ratchet-like rotations of a laterally constricted SkL.[98] Whereas, for lateral thermal gradients, unconfined skyrmions were found to move while experiencing the skyrmion Hall effect.[99] Although exciting, thermal effects may be difficult to use practically.

Secondly, in the case of the biskyrmion-hosting La$_{1.37}$Sr$_{1.63}$Mn$_2$O$_7$ system, current-based STTs generated by low $J_C$ ~$10^7$ A/m$^2$ triggered the lateral motion of SkLs.[36] This STT effect emerges from spin-polarized charge current transferring a net angular momentum to the localized spins, effectively resulting in skyrmion motion. Moreover, the low $J_C$ is likely a result of high crystalline quality of the oxide samples minimizing defect pinning.[16]

Subsequently, although SOT driven motion of skyrmions is yet to be directly confirmed in oxides, recent studies have used SOTs to displace chiral FiM DWs. Particularly, in Pt/garnet-ferrite heterostructures,[100] lateral charge currents in the Pt layer resulted in pure spin currents normal to the interface, due to spin Hall effect.[101] These spin currents exerts a damping-like SOT and push the chiral DWs into motion, with ultrafast speeds on the order ~km/s, thanks to the low Gilbert damping.[72] At these speeds, magnetic textures start to reach the speed-limit of magnons thereby reproducing spintronic analogues of relativistic physics (e.g., Lorentz contraction). It is expected that similar processes should be able to drive skyrmions and other topological textures in an ultrafast manner via SOTs in Pt-capped AFM oxides.[71]

Meanwhile, for detection of topological textures, the TNcMR effect has emerged to be very promising. It is electrically sensitive to the magnetic textures through local variations in the electronic density of states (DOS). For instance, a collinear magnetic background (representing bit value '0') would register a larger tunnelling current in comparison to the case when a topological texture (representing '1') is present. This effect can be harnessed by integrating tunnel junctions onto topological racetracks making them scalable. Thus far, the TNcMR effect has been realized only in magnetic metallic heterostructures via ultrasharp STM tips.[26, 102] Moreover, its experimental demonstration in skyrmion hosting oxides is also pending. Nonetheless, in strongly correlated metallic oxides with d-electron dominated conduction the electronic bands are usually narrow in the energy spread with large DOS, which is closely related to the local magnetic arrangement. Hence, one could expect the TNcMR effect to be enhanced in the case of correlated oxides. To detect textures in insulating AFM oxides, one could harness interfacial exchange interactions to couple the topological AFM order to an adjacent magnetic metal,[82, 86] which could in turn be read-out via TNcMR.

## 4. Outlook

Given that a wide variety of topological textures and control pathways have already been demonstrated in oxides, the stage is now set to explore oxide skyrmionics. Imminent steps are to stabilize topological textures in FiM or AFM oxides that are coupled to a ferroelectric order, enabling non-volatile control over the topological stability landscape, at room temperature and above. This can be facilitated by careful heterostructure designing of emergent phenomena.

Subsequently, it is important to harness current-based SOTs to realize ultrafast motion of textures along racetracks. Finally, incorporation of tunnel junctions to detect topological textures via TNcMR will allow scalable read out of bits. Integrating all these aspects in a *single device* would enable practical exploitation of oxides for low-power multifunctional skyrmionics.

## Acknowledgments

This work was supported by the Agency for Science, Technology and Research under the AME Individual Research Grant (A1983c0034) and the National Research Foundation (Singapore) under the Competitive Research Program (NRF-CRP15-2015-01).

**Declarations**

The authors have no competing interests to declare that are relevant to the content of this article.